\documentclass[letterpaper,twocolumn,10pt]{article}
\usepackage{./commons/usenix-2020-09}
\usepackage{CJKutf8}
\usepackage{amsmath}
\usepackage{graphicx}
\usepackage[tight,footnotesize]{subfigure}
\usepackage{multirow}
\usepackage{fancyvrb}
\usepackage[flushleft, online]{threeparttable}
\usepackage{cite}
 \usepackage{xurl}
\usepackage{array}
\usepackage{algorithmic}
\usepackage{booktabs}
\usepackage{xcolor}
\usepackage{ulem}
\usepackage{authblk}

\newcommand{\subject}[1]{\vspace{3pt}\noindent\textbf{#1}}
\newcommand{\subsubject}[1]{\vspace{3pt}\noindent\textit{#1}}

\newcommand\malurl[1]{\href{notalink}{{\nolinkurl{#1}}}}
\newcounter{finding}
\newcommand{\finding}[1]{
\vspace{3pt}
\noindent
\framebox{
\begin{minipage}[b]{3.2in}
\noindent \textbf{Finding \Roman{finding}}: \textit{#1}
\stepcounter{finding}
\end{minipage}
}
\vspace{-5pt}
}

\newcommand{\ignore}[1]{}
\newcommand{\cut}[1]{}

\title{An Empirical Study of Bandwidth Sharing Services}
\title{Towards Analyzing and Detecting Network Traffic of Residential Proxies}
\title{Shining Light into the Tunnel: Understanding and Classifying Network Traffic of Residential Proxies}

\begin{document}
\author[1]{Ronghong Huang}
\author[2]{Dongfang Zhao}
\author[1]{Xianghang Mi\thanks{Corresponding author at xmi@ustc.edu.cn}}
\author[2]{Xiaofeng Wang}

\affil[1]{University of Science and Technology of China}
\affil[2]{Indiana University Bloomington}

\affil[ ]{\url{https://chasesecurity.github.io/bandwidth_sharing}}


\maketitle

\begin{abstract}
    Emerging in recent years, residential proxies (RESIPs) feature multiple unique characteristics when compared with traditional network proxies (e.g., commercial VPNs), particularly, the deployment in residential networks rather than data center networks, the worldwide distribution in tens of thousands of cities and ISPs, and the large scale of millions of exit nodes. All these factors allow RESIP users to effectively masquerade their traffic flows as ones from authentic residential users, which leads to the increasing adoption of RESIP services, especially in malicious online activities. However, regarding the (malicious) usage of RESIPs (i.e., what traffic is relayed by RESIPs), current understanding turns out to be insufficient. Particularly,  previous works on RESIP traffic studied only the maliciousness of web traffic destinations and the suspicious patterns of visiting popular websites. Also, a general methodology is missing regarding capturing large-scale RESIP traffic and analyzing RESIP traffic for security risks.  Furthermore, considering many RESIP nodes are found to be located in corporate networks and are deployed without proper authorization from device owners or network administrators, it is becoming increasingly necessary to detect and block RESIP traffic flows, which unfortunately is impeded by the scarcity of realistic RESIP traffic datasets and effective detection methodologies.

To fill in these gaps, multiple novel tools have been designed and implemented in this study, which include a general framework to deploy RESIP nodes and collect RESIP traffic in a distributed manner, a RESIP traffic analyzer to efficiently process RESIP traffic logs and surface out suspicious traffic flows, and multiple machine learning based RESIP traffic classifiers to timely and accurately detect whether a given traffic flow is RESIP traffic or not. 
As the results, we have collected and will release the largest-ever and realistic RESIP traffic dataset, which is of 3TB in size, consists of over 116 million traffic flows, and involves traffic destinations of 188K unique IP addresses. Also, leveraging the RESIP traffic analyzer, multiple novel security findings have been distilled regarding the malicious usage of RESIPs, e.g., relaying large-scale email spam activities targeting millions of recipients, and the masquerade of miscreants as local residents when suspiciously visiting sensitive websites operated by critical organizations, e.g., governments,  military agencies, and companies operating critical infrastructures. Lastly, our machine learning based RESIP traffic classifiers turn out to be both effective and efficient. Particularly, when classifying whether a traffic flow is relayed by RESIPs or not, our transformer-based classifier has achieved a recall of 93.04\% and a precision of 92.87\%, by only ingesting the first 5 packets of each given traffic flow.
\end{abstract}

\section{Introduction}
\label{sec:intro}

Residential proxies (RESIPs) are web proxies located in residential or cellular networks. Compared to traditional web proxies (e.g.,  VPNs and open web proxies), RESIPs tend to have a much larger scale as well as a wider distribution, as revealed by previous studies~\cite{mi2019resident, DBLP:conf/ndss/MiTLLQ021, DBLP:conf/ccs/YangYMTGLZD22}. 
Given such a large scale, previous studies have investigated how RESIPs are recruited by RESIP services, which leads to the discovery of multiple controversial or even malicious recruitment channels, e.g., installing potentially unwanted programs (PUPs) on Windows devices~\cite{mi2019resident}, distributing mobile proxy apps with ambiguous user consent prompts~\cite{DBLP:conf/ndss/MiTLLQ021}, etc. 

Furthermore, research efforts have also been invested to understand the usage (abuse) of RESIPs. Leveraging traffic logs of PUPs that have been confirmed to serve as RESIPs,  Mi, et al.~\cite{mi2019resident} discovered malicious RESIP traffic of multiple categories, such as advertisement frauds and blackhat search engine optimization (SEO). However, it is unclear whether the PUP traffic logs are exclusively RESIP traffic, since these PUPs can have other malicious functionalities in the meantime, which may contribute to the resulting malicious traffic logs.
Besides, RESIP traffic relayed by mobile proxy apps had also been captured and analyzed~\cite{DBLP:conf/ndss/MiTLLQ021}, from which, suspicious traffic of similar categories got observed. However, the authors didn't specify the scale of the captured RESIP traffic, nor do they release the RESIP traffic to the public. Given the abuse of RESIPs in various malicious activities, efforts have been further invested to detect RESIP traffic. Some studies~\cite{resident_flow_detection_ml, chiapponi_BADPASS_2022} targeted traffic flows relayed by RESIPs, namely \textit{the relayed flows},  while others~\cite{DBLP:conf/ndss/MiTLLQ021} focused on the tunnel connections between a RESIP node and a RESIP gateway server, namely \textit{the tunnel flows}. However, none of these methodologies has been evaluated on wild RESIP traffic.

Stepping forward from these previous studies, this paper has made contributions in several aspects. First of all, we have collected the largest-ever dataset of over 3TB RESIP traffic, which consists of over 116 million traffic flows towards traffic destinations of 188K IP addresses and 116K fully qualified domain names. Furthermore, this dataset of wild RESIP traffic gives us the opportunity to explore the feasibility of applying machine learning to the detection of both relayed flows and tunnel flows. As a result, multiple machine learning classifiers have been built up along with very good performance achieved. Particularly,  our relayed flow classifier has achieved a recall of 88.80\% and a precision of 96.20\%, while the performance for our tunnel flow classifier is 91.39\% in recall and 95.71\% in precision. Furthermore, an in-depth analysis has been carried out for the traffic flows relayed by RESIPs, which has distilled multiple novel findings regarding how RESIPs are being abused in malicious activities.

Our study would not be possible without the design and implementation of a set of novel methodologies. The first challenge we encountered is how to collect RESIP traffic at scale, especially considering RESIP services are known to keep their RESIPs as black boxes. We observe that some RESIP providers also operate bandwidth sharing (BS) as a service in the meantime. And the representative BS providers under our study include PacketStream~\cite{packetstream}, IPRoyal~\cite{iproyal}, and Honeygain~\cite{honeygain}. A BS service is offered to monetize the idle bandwidth available on a BS node (e.g., a laptop device), and the node operator will be paid proportionally to the volume of bandwidth the node has contributed. Through preliminary experiments of running BS nodes and analyzing the resulting traffic, we have confirmed that the bandwidth provided by BS nodes is exclusively used to relay RESIP traffic. In other words, BS nodes are exclusively used as RESIP nodes. Therefore, we collected RESIP traffic by deploying BS nodes at scale, during which, we also addressed other challenges, such as how to trigger effective BS (RESIP) traffic, and how to deploy BS nodes and collect traffic in a scalable way, which jointly lead to a general RESIP traffic collector.
Then, a RESIP traffic analyzer is built up to efficiently process RESIP traffic and effectively locate the inherent security risks. And our in-depth analysis of RESIP traffic  has distilled a set of novel and concerning security findings as highlighted soon later. 
Considering the concerning security risks of RESIP traffic,  we have further designed a set of robust features and built up multiple machine learning classifiers which enable early-time detection of both relayed flows and tunnel flows with high precision. 

Below, we summarize the key findings and contributions of this study. 

First of all, we observe, for the first time, that RESIPs have been extensively used to masquerade attackers as local residents when visiting sensitive websites operated by government agencies, military agencies, and organizations in other public sectors. In total, we have observed suspicious traffic flows toward 73 different security-sensitive websites, which belong to diverse categories, e.g., industrial control systems, airport control systems, watering control systems, remote desktops inside government agencies, and various military websites. Among these websites, 31\% are operated by governments, 18\% are for military agencies, and another 21\% are for other organizations in the public sector, e.g., hospitals, education agencies, railway management companies, and research agencies. Also, a strong correlation is observed between the country of our RESIP nodes and the country wherein these sensitive websites are hosted.

Besides,  still for the first time, we confirm that RESIPs have also been used to relay a large volume of email spamming activities and suspicious email retrieving activities. Particularly, email spamming activities involve 465K sender email addresses, 2.29 million recipient email addresses, and 549K unique spam messages. What is also observed is various evasion techniques as adopted by email spammers, e.g., text paraphrasing,  complicated combinations of character encodings, email encodings, homoglyphs, etc.

Furthermore, to enable early-time detection of RESIP traffic, a set of machine learning classifiers have been built up along with good performance demonstrated. Particularly, our transformer-based classifiers not only remove the burden of manual feature engineering, but also achieve a high detection performance by only ingesting the first 5 packets of each traffic flow. For instance, the transformer-based model for relayed traffic classification has achieved a recall of 93.04\% and a precision of 92.87\%. On the other hand, our feature-based models have better explainability as well as decent performance, e.g., when considering only the first 4 upstream packets, our feature-based model for tunnel flow classification can achieve a recall of 91.94\% and a precision of 94.87\%. To the best of our knowledge, this is the first work which applies machine learning to the detection of RESIP traffic. 

Our contributions can be summarized below. 

\vspace{2pt}\noindent$\bullet$ We have designed and implemented a novel methodology to collect RESIP traffic at a large scale and to analyze RESIP traffic for security risks.

\vspace{2pt}\noindent$\bullet$ Leveraging the RESIP traffic collector, we have collected the largest-ever dataset of wild RESIP traffic.
This traffic dataset will be available upon request, so as to foster future research while avoiding potential security or privacy risks.

\vspace{2pt}\noindent$\bullet$ Leveraging the RESIP traffic analyzer, we have identified multiple novel and concerning security findings regarding the malicious usage of RESIPs. 

\vspace{2pt}\noindent$\bullet$ We have demonstrated,  for the first time, the feasibility of applying machine learning to the detection of RESIP traffic. 

\section{Background and Related Works}
\label{sec:background}
\begin{figure}
    \centering
    \includegraphics[width=.9\columnwidth]{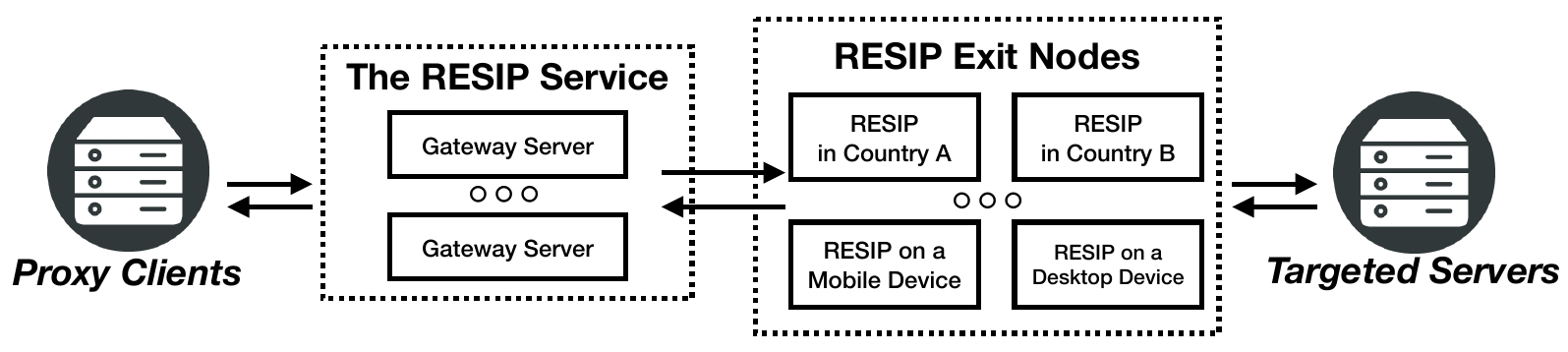}
    \caption{The backconnect proxy mode of RESIPs.}
    \label{fig:resip_proxy_mode}
\end{figure}

\subject{Residential proxies}. Residential proxies (RESIP) are web proxies located in either residential or cellular networks. As illustrated in Figure~\ref{fig:resip_proxy_mode}, a typical RESIP works in a backconnect mode wherein proxy traffic sourced from a proxy customer will be first sent to the gateway server, which in turn forwards the traffic to a RESIP node, before exiting to the traffic destination. 
What's more, most RESIP services will allow proxy customers to specify where they want to exit the relayed traffic, in terms of countries and cities. Also, proxy customers are able to stick their traffic to the same exit node, either through passing the same session id or connecting to some sticky proxy gateways which usually bind to an exit node for every 5 or 10 minutes~\cite{proxyrackstickyip}. 
Varied by RESIP services, multiple proxy protocols can be supported, which range from HTTP/HTTPS and SOCKS to a variety of VPN protocols, e.g., PPTP, L2TP, and OpenVPN. Regarding pricing, a RESIP service usually offers several monthly subscription options which differ in price and resource constraints (e.g., traffic volume). In our study, the terms of \textit{RESIP} and \textit{RESIP IP} will be used interchangeably since the IP address is the only publicly available information we can have to refer to a RESIP.  


As aforementioned, three RESIP/BS services are chosen in our study to understand RESIP traffic and experiment different methodologies for RESIP traffic classification. These RESIP services include PacketStream,
IPRoyal, and Honeygain, which, as detailed later in \S\ref{sec:ecosystem}, share with RESIP services studied in previous works~\cite{mi2019resident, DBLP:conf/ndss/MiTLLQ021, DBLP:conf/ccs/YangYMTGLZD22},  a set of common characteristics including service model, exit node recruitment, traffic pattern/protocols, and usage. We thus believe our measurement results and methodologies can be generalizable to many other RESIP services. 

A set of works focus on profiling the scale and distribution of residential proxies (RESIPs)~\cite{mi2019resident, DBLP:conf/ndss/MiTLLQ021,DBLP:conf/ccs/YangYMTGLZD22}. Particularly, Mi, et al.~\cite{mi2019resident} captured more than 6 million residential proxies during 2017 for 5 RESIP services, while Yang et al.~\cite{DBLP:conf/ccs/YangYMTGLZD22} moved forward to profile RESIPs in China along with over 4 million China RESIPs captured during 2021.
Besides, Chiapponi et al~\cite{DBLP:conf/icsoc/ChiapponiDTCT20} proposed several mathematical approaches to model the scale of RESIPs. 
Furthermore, several recruitment channels for RESIPs~\cite{upnproxy, mi2019resident, DBLP:conf/ndss/MiTLLQ021} have been revealed, e.g., the recruitment of potentially unwanted programs (PUPs)~\cite{mi2019resident}, and deploying RESIP nodes as 3rd-party SDKs of mobile apps~\cite{DBLP:conf/ndss/MiTLLQ021}. Besides, some studies also profiled the usage of RESIPs in malicious activities. Particularly, Mi, et al.~\cite{mi2019resident} found out that RESIPs had been abused to relay malicious traffic towards popular websites such as advertisement frauds and blackhat search engine optimization (SEO). Compared with these previous studies, our study has moved forward in several aspects including the novel methodology to collect and analyze RESIP traffic, the collection of a large-scale RESIP traffic dataset, and the discovery of previously unknown malicious activities as relayed by RESIPs.

Considering the abuse of RESIPs in various malicious activities, multiple studies~\cite{DBLP:conf/ndss/MiTLLQ021,chiapponi_BADPASS_2022} explored possible features to identify RESIP traffic flows and designed various threshold-based methods to distinguish RESIP flows from non-RESIP ones. 
Moving forward from these studies, our study has explored and demonstrated, for the first time,  the feasibility of machine learning based detection of RESIP traffic.

\subject{Encrypted traffic classification}.  Compared with plaintext traffic classification, classifying encrypted network traffic incurs more challenges along with less robust features. Draper-Gil et al.\cite{draper2016characterization} explored time-related features, e.g., the inter-packet arrival time, while Wang, et al.~\cite{wang2017end, wang2017hast} demonstrated the effectiveness of one-dimensional CNN on encrypted traffic classification.
Furthermore, He, et al.~\cite{he2020pert} adopted the spirit of word embedding, explored generic traffic representation, and proposed PERT. 
Moving forward, Lin, et al.~\cite{lin2022bert} proposed and evaluated Encrypted Traffic Bidirectional Encoder Representations from Transformer (ET-BERT). ET-BERT considers not only the inter-packet correlation but also the correlation between the incoming and outgoing flows of the same traffic session. Applying ET-BERT to multiple tasks of encrypted traffic classification  has achieved state-of-the-art performance. In this study, we have fine-tuned the ET-BERT model and built up multiple transformer-based RESIP traffic classifiers. 
Besides, Jorgensen et al.~\cite{jorgensen2022extensible} observed several limitations of the ISCXVPN2016 datasets (e.g.,  some VPN-labelled packets were found to have unencrypted payloads), and released a new traffic dataset under the name of \textit{VNAT}. In our study, both the ISCXVPN2016 and VNAT are used for RESIP traffic classification (\S\ref{sec:defense}).

\section{ Collecting and Analyzing RESIP Traffic}
\label{sec:method}
\begin{figure}
    \centering
    \includegraphics[width=.9\columnwidth]{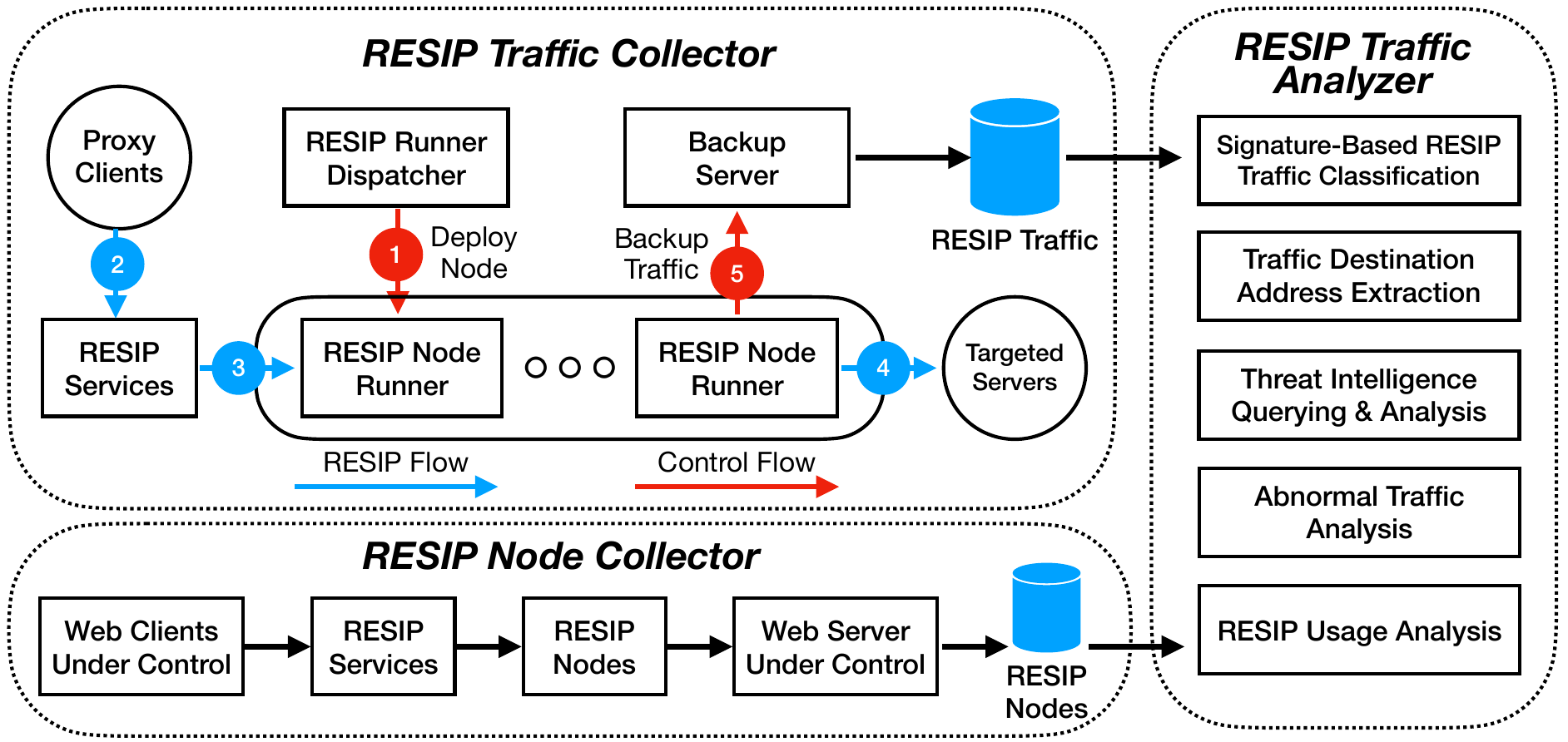}
    \caption{The methodology to collect and analyze RESIP traffic.}
    \label{fig:method}
\end{figure}
In this section, we move to present the methodologies for collecting and understanding wild RESIP traffic. 
As illustrated in Figure~\ref{fig:method},   
a RESIP traffic collector is first designed and deployed to run RESIP nodes across locations and collect wild and real-world RESIP traffic, which is detailed in \S\ref{subsec:method_bs_traffic}.  Also, to further profile the scale of the selected RESIP services and confirm that RESIP nodes under our deployment are indeed used to relay RESIP traffic, a RESIP node collector (\S\ref{subsec:method_resip_nodes}) is built up to collect public IP addresses of RESIPs. Lastly, given the collected RESIP traffic, to gain an in-depth understanding of the involved activities and security risks, an analysis toolchain is built up under the name of \textit{the RESIP traffic analyzer} (\S\ref{subsec:method_traffic_analyzer}). 

\subsection{The RESIP Traffic Collector}
\label{subsec:method_bs_traffic}
\subject{Design.} As revealed by our preliminary experiments of deploying bandwidth-sharing (BS) nodes, BS nodes are confirmed to be exclusively used as RESIPs to relay traffic. Therefore, if not especially noted, we will use BS and RESIP interchangeably in the rest of this paper. 
To collect RESIP traffic, a RESIP traffic collector is built up, which consists of two major components: a set of RESIP node runners and a central RESIP node dispatcher (i.e., controller). 
Specifically, the dispatcher is designed to instruct the RESIP node runners regarding how to deploy RESIP nodes and how to back up the resulting network traffic along with other log files.

On the other hand, a RESIP node runner is responsible for running a RESIP node for a period of time by following deployment instructions from the central dispatcher. Besides, as some runner machines may be subject to dynamic IP assignments and logging their changing IP addresses can facilitate further data correlation with RESIP IP dataset (as detailed in \S\ref{subsec:method_resip_nodes}), each runner will also run a side job to query and log the runner machine's latest public IP addresses, which is achieved through periodically querying IPinfo's API~\footnote{https://ipinfo.io/json}. 

\subject{Implementation and Deployment.} Although the three RESIP providers (PacketStream, Honeygain, and IPRoyal Pawns) all offer docker containers to recruit RESIP nodes, it turns out to be non-trivial to successfully trigger effective RESIP traffic. We started by executing RESIP nodes in cloud computing platforms, but all received error messages regardless of the RESIP providers, saying that the IPs of the nodes were not qualified. Further investigation shows that RESIP providers require either implicitly or explicitly that the RESIP nodes should be deployed on residential or cellular IPs\footnote{https://support.honeygain.com/hc/en-us/articles/360011078760-Error-Unusable-network}. Therefore, once a RESIP node is initiated, it will first check with the RESIP gateway servers to decide whether its public IP address is residential or not. If not, the node container will exit with an error message suggesting the current network is not usable. To address this problem, for runners in the USA, we deployed them in servers located in a university institution whose networks are considered valid by all three RESIP providers. For workers in China, we tried all publicly available cloud computing options and found residential virtual private servers (VPS) can partially work.  Residential VPSes are offered in China by cloud computing services which claim that these VPSes are deployed in residential networks. By deploying RESIP nodes on these residential VPSes in China, we have successfully triggered effective traffic for both PacketStream and Honeygain but still failed for IPRoyal. 

When deploying the RESIP traffic collector, RESIP node runners were deployed in both the USA and China, in an attempt to profile the impact of location on RESIP traffic. In total, The deployment consists of 2 runners in the US for each of the three RESIP providers, 8 runners in China for PacketStream, 8 runners for Honeygain in China, and one central server deployed as both the dispatcher and the backup server. 

%
\subsection{The RESIP Node Collector}
\label{subsec:method_resip_nodes}
To further confirm that the RESIP nodes under our deployment have indeed been used to relay RESIP traffic as well as evaluate the scale of aforementioned RESIP services,  we move to capture RESIP nodes (i.e., RESIP IP addresses). 
As revealed in previous works~\cite{mi2019resident}, a residential proxy works in backconnect mode wherein the exit RESIP node is hidden behind the proxy gateway. To capture residential proxies, we adopted the RESIP node collection framework as proposed in \cite{mi2019resident}. As illustrated in Figure~\ref{fig:method}, to capture RESIP nodes,  web clients and web servers are first deployed under our control. Then,  probes in the form of simple HTTP requests will be sent from web clients, forwarded through proxy gateway servers, exited at RESIP nodes, before reaching the web servers. Through this process, the web servers will be able to observe the public IP address of each involved RESIP node.

The first step of the infiltration is to get access to the RESIP services. And we successfully made it for PacketStream and IPRoyal but failed for Honeygain due to its strict background checking. We then conducted the infiltration during the period of RESIP traffic collection, for both PacketStream and IPRoyal, which resulted in 58,508,588 probes being successfully sent and 2,122,255 unique RESIP IP addresses captured, for which, more details will be presented in \S\ref{subsec:bs_nodes}. 

\subsection{The RESIP Traffic Analyzer}
\label{subsec:method_traffic_analyzer}
The large scale of the captured RESIP traffic renders manual analysis impractical. Therefore, as illustrated in Figure~\ref{fig:method}, a RESIP traffic analyzer consisting of multiple traffic processing modules was designed and implemented. Below, we introduce these tools one by one with a focus on their design goals and functionalities. 

\subject{Signature-Based RESIP Traffic Classification}. One thing to note, not all the captured RESIP traffic flows are ones that are sourced from proxy customers. Instead, the collected RESIP traffic flows can be divided into three groups. One group comprises the flows between the RESIP nodes and the remote proxy gateway servers. These flows are set up to accept incoming relaying requests from proxy servers, and forward response traffic from traffic destinations back to the proxy gateway servers. We thus name them as \textit{tunnel flows}.
The second group is the flows used by RESIPs to communicate with the traffic destinations. And we name such flows as \textit{relayed flows}, which are the \textit{true} traffic relayed by RESIPs. The third group consists of control-plane flows between a RESIP node and its respective RESIP service, e.g., flows to sync node stats, and connections to pull the list of proxy gateway servers. And we name this group as \textit{the control flows}. 

To get a deep understanding of the underlying relaying activities, it is necessary to decide whether a given network connection is a relayed flow or not. We pursue this task through a signature-based classifier. 
Specifically, we observe that both control flows and tunnel flows have either service-specific domain names or customized protocols (e.g., customized TLS), regardless of the RESIP service. Therefore, these domain names and customized protocol patterns can serve as robust signatures to extract the respective flows while the left ones can be reliably considered as relayed flows towards traffic destinations. For instance, the domain signature for PacketStream's tunnel flows is \textit{proxy.packetstream.io} while  \textit{api.iproyal.com} serves as a domain signature for control flows of IPRoyal.  Also, IPRoyal adopts a customized TLS protocol for its tunnel flows, which makes IPRoyal's tunnel flows distinguishable from others in that there is no TLS handshake stage but the flow payloads can be recognized as TLS records. Leveraging both domain signatures and protocol signatures (in the case of IPRoyal's tunnel flows), a signature-based RESIP traffic classifier is built up to decide whether a RESIP flow is a control flow,  a tunnel flow,  or a relayed flow.




\subject{Traffic destination addresses extraction}. 
To facilitate statistical analysis of relayed RESIP traffic, the next step is to extract all the traffic destinations and aggregate statistical attributes around these destination addresses. Specifically, for each given relayed flow, we first extract all its destination addresses including the IP address, the TCP or UDP port, the fully qualified domain name (FQDN) if available, as well as the URL if any. 
Among these addresses, the IP address and the transport-layer port can be easily extracted from the packets. However, none of them can reveal much human-readable information regarding for what activities the flow is used. Instead, the fully qualified domain name (FQDN) to which a traffic flow is intended to visit can be more helpful in terms of understanding its activities. 
Therefore, we have designed a methodology to accurately associate a relayed traffic flow with the domain name that it is used to visit. Specifically, for HTTP traffic, the domain name can be easily extracted from the plaintext payload. Furthermore, for HTTPS connections, we first try to extract the domain name from the server name indicator (SNI) extension of TLS. For HTTPS connections which don't have the SNI extension or other non-HTTP connections, we first build up mappings between domains and IPs, through parsing DNS queries and responses recorded in the same PCAP file. Then, a TCP flow will be considered to have the domain name identified only when its remote IP address can be uniquely mapped to a domain name.  


\subject{Querying and analyzing threat intelligence of traffic destinations.} Given the hierarchical destination addresses, to profile their maliciousness, we also analyzed threat reports from multiple representative threat intelligence datasets: VirusTotal, IBM X-Force, and URLhaus. Adapting the availability of threat reports for different network addresses, three types of destination addresses were queried, which include the IP address, the FQDN, and the URL.

\subject{Abnormal traffic analysis}. Furthermore, a set of criteria has been defined to surface abnormal or suspicious traffic flows and their destination addresses. And a traffic flow will be considered as abnormal and deserves further manual investigation when it satisfies one of the following criteria: 1) first of all, its destination addresses receive one or more alerts from aforementioned threat reports; 2) its destination ports are not commonly used ones; 3) it has no FQDN identified; 4) its application layer protocol doesn't belong to widespread web protocols including HTTP, HTTPS, and QUIC; 5) Not all its destination addresses have been observed in China RESIP nodes and US RESIP nodes. Given relayed flows matching these rules, in-depth manual analysis will be followed, through which, we have identified a set of novel findings, e.g., email spamming activities, and visiting security-sensitive Chinese websites through RESIP nodes located in China. More details will be presented later in \S\ref{sec:sec_risks}. 


\subject{RESIP usage analysis.}
As aforementioned, we concurrently collected active RESIP nodes through infiltration while collecting RESIP traffic. If RESIP nodes under our control were indeed used to relay traffic and our infiltration had good coverage of RESIP nodes, it is very likely that we can observe some infiltration traffic flows in the collected RESIP traffic.  We answer this question by conducting direct correlation analysis between the RESIP infiltration dataset and the RESIP traffic dataset. Specifically, a RESIP node is considered to have been used to relay the infiltration traffic only when one or more pairs of traffic flows from the two datasets are accurately matched in terms of the traffic destination, the timestamp, and the traffic size.
\subsection{Ethical Considerations and Limitations}
\label{subsec:method_ethics}
\subject{Ethical considerations}. 
Our study has gone through a thorough IRB review and is deemed by the IRB reviewers as exempt from an approval.
Throughout our study, we place a strong emphasis on ethics and have carefully designed and scrutinized our methodologies and results, so as to minimize any potential ethical issues. Particularly, when collecting RESIP traffic and RESIP nodes, the raw datasets are stored on our secure research servers with restricted access, and any data transmission has been encrypted to prevent unauthorized access. Furthermore, when manually analyzing the raw network traffic, our primary focus is on understanding the traffic patterns, summarizing the traffic categories, and locating security-relevant issues. And our researchers are explicitly instructed to disregard any personally identifiable information (PII) if revealed in unencrypted traffic. Additionally, with respect to the presentation of our measurement results, we ensure that only statistical data points are generated and disclosed. Specific cases are provided only when they have been thoroughly vetted and confirmed to be free from privacy risks.

Also, as detailed in \S\ref{sec:sec_risks}, RESIP nodes were found to have relayed malicious traffic such as Email spamming. However, due to factors illustrated below, we were not able to block such malicious traffic. Specifically, when deploying RESIP nodes and collecting their traffic, we didn't know in advance regarding what (malicious) traffic would be relayed by RESIP nodes, not to mention how to block them.
Instead, the findings of malicious traffic were distilled when analyzing the resulting RESIP traffic. At that time point, our traffic collection jobs had already been finished. 
However, our measurement results highlight the necessity of enforcing preventive measures when deploying RESIP nodes and capturing RESIP traffic in the future. Example preventive measures can be the blocking of any Email traffic or traffic towards unrecognized TCP/UDP ports.  

\subject{Limitations}. Our methodologies still suffer from a set of limitations. Particularly, when collecting RESIP traffic, our RESIP nodes were constrained to two locations, one in the USA and the other in China, and we may miss traffic with a preference for RESIPs in other locations, e.g., Europe. 
Furthermore, regarding our signature-based RESIP traffic classifier, we focus on efficiently distinguishing RESIP traffic for the three services under our study, and the classifier will not generalize to any unknown RESIP service, unless we know the respective signatures. Stepping forward, a set of general RESIP traffic classifiers have also been built up as a defensive measure, as detailed later in \S\ref{sec:defense}.
\section{The RESIP Ecosystem}
\label{sec:ecosystem}
In this section, we present an up-to-date characterization of the RESIP ecosystem, which has distilled multiple novel observations when compared to previous studies. We first present how a RESIP service is operated from the technical perspective as well as the scale of these services in terms of RESIP nodes, as detailed in \S\ref{subsec:bs_service} and \S\ref{subsec:bs_nodes}. Then, in \S\ref{subsec:bs_traffic}, we move the spotlight to RESIP traffic with a focus on the scale, traffic patterns, and traffic categories.

\subsection{RESIP Services}
\label{subsec:bs_service}

\subject{Protocols}. Through running BS nodes and observing their network traffic, we conclude that BS nodes are used to serve residential proxies and relay third-party network traffic. Therefore, we specify here how a BS node (a RESIP) interacts with different parties to fulfill this functionality. 
In general, to relay traffic, a RESIP node will initiate and maintain one or more persistent tunnel connections with the RESIP gateway server, and the tunnel connection will be used to receive relaying traffic from the proxy server before forwarding it to the traffic destination, and vice versa. Also, regardless of the RESIP services, the tunnel connection between the RESIP node and the RESIP server is encrypted.

However, several discrepancies in the RESIP protocol still exist among BS services. Particularly, regarding how a BS node learns the addresses of the proxy gateway server, RESIP nodes in PacketStream are pre-configured with the proxy server's domain name, namely, \textit{proxy.packetstream.io}, while BS nodes of either IPRoyal or Honeygain will query a pre-configured control server (e.g., api.honeygain.com in the case of Honeygain or api.iproyal.com for IPRoyal) to extract the latest list of proxy gateway servers before setting up the tunnel connections. Then, when connecting to the proxy server, PacketStream and Honeygain adopt the standard HTTPS protocol while IPRoyal deploys a customized encrypted protocol. 



\finding{ 
DNS over HTTPS (DoH) has been adopted by one of the three RESIP services, namely Honeygain, thus enhancing the privacy and anonymity of RESIP traffic.
}

Furthermore, when relaying traffic towards a domain name, another discrepancy occurs when querying DNS records. Specifically, nodes of PacketStrem and IPRoyal utilize the traditional UDP-based DNS protocol wherein both DNS queries and responses are in plaintext. Also, the queries will be sent to the DNS resolver as configured by the local network. On the other hand,  a Honeygain node adopts the protocol of DNS over HTTPS (DoH), a more privacy-preserving protocol to query DNS servers over HTTPS. Specifically, a Honeygain node will send all DNS queries to https://cloudflare-dns.com/dns-query.



\subject{The pricing policy}. The pricing policy varies across services. In PacketStream,  by running a RESIP node, the node operator can earn 0.1\$ for every 1GB of traffic relayed through the node. On the other side, a proxy user, i.e., a bandwidth consumer, will be charged 1\$/GB for relaying traffic through RESIP nodes. 
For IPRoyal, the price for bandwidth providers is 0.2\$/GB, while it used to be 0.8\$/GB in Feb 2023 for bandwidth consumers and increased to  \$7/GB in Jul 2023.
Honeygain's pricing policy for bandwidth providers is more complicated, and it depends on the available bandwidth and the active duration of a given node. For instance, a bandwidth provider can earn 46\$ per month for sharing 5GB traffic per day and being active for 4 hours per day. Since we failed to gain access to their residential proxies, it is unclear regarding their pricing policies for proxy users.

\begin{table}
    \centering
    \footnotesize
    \caption{The scale and distribution of RESIP nodes.}
    \label{tab:stat_bs_node}
    \begin{tabular}{cllll}
    \toprule
       Provider & IPs & /24 IPv4 & /16 IPv4 & Countries  \\
       \midrule
       PacketStream &  300,723 & 136,930 & 17,754 &  185\\
        
        IPRoyal & 1,915,971 & 481,184 & 23,910 &  213\\
        All & 2,122,255 & 525,403 & 24,761 &  213\\
        \bottomrule
    \end{tabular}
\end{table}

\subsection{RESIP Nodes}
\label{subsec:bs_nodes}
As detailed in \S\ref{subsec:method_resip_nodes}, we captured RESIP nodes through subscribing to RESIP services offered by PacketStream and IPRoyal, and relaying probing traffic through RESIP nodes to web servers under our control. 
As detailed below, this allows us to profile the scale of RESIP services, the global distribution of their RESIP nodes, as well as further demonstrate that the RESIP nodes under our control have been indeed used to relay network traffic.

\finding{Along with large-scale and globally distributed RESIP nodes, the RESIP services under our study are representative players in the RESIP ecosystem.}

\begin{figure}
    \centering
    \includegraphics[width=.8\columnwidth]{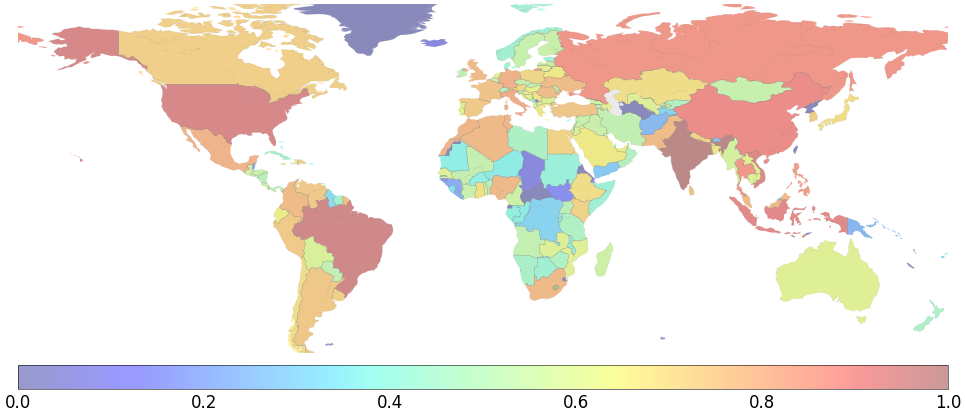}
    \caption{The distribution of RESIP nodes in a world heatmap.}
    \label{fig:node_dist}
\end{figure}

\subject{Scale.} In total, we have captured 2,122,255 RESIP IP addresses through 58 million probes for PacketStream and IPRoyal. We queried IPinfo for the geolocation information of IP addresses, e.g., the country, city, and ISP.  These RESIP nodes are widely distributed in 213 countries and regions, as illustrated in Table~\ref{tab:stat_bs_node} and the world heatmap of Figure~\ref{fig:node_dist}. Also, top 10 countries contribute 51.16\% RESIP nodes, e.g., India accounts for 11.52\%, Brazil contributes 7.32\%, and it is 6.41\%  for the United States.

\subject{RESIP nodes under our control}. All RESIP nodes under our control have been observed in relaying our infiltration probes. In total, our RESIP nodes, accounting for only 0.0005\% of over 2 million RESIP nodes, have relayed 2,584 infiltration probes (0.0045\% of all probes), which suggests the high coverage of our infiltration process for RESIP nodes. Also, this finding strengthens our observation that bandwidth-sharing nodes are exclusively used as RESIP nodes. 


\subject{A comparison with previous RESIP IP datasets}. We also compare our RESIP node dataset with residential proxy datasets as released by previous works. These datasets include  \textit{RESIP-2017} collected during 2017~\cite{mi2019resident}, \textit{RESIP-2019} in 2019~\cite{DBLP:conf/ndss/MiTLLQ021}, and \textit{RESIP-2021} in 2021~\cite{DBLP:conf/ccs/YangYMTGLZD22}. Among the 2.12 million RESIP IPs observed in our study, only 251,898 (11.87\%) were observed in one or more of these previous residential proxy datasets, while it is 12.10\% for IPRoyal and 11.73\% for PacketStream separately. 
Multiple factors can contribute to this low intersection rate. One is that RESIP nodes can migrate quickly across network blocks due to either device movement or the dynamic IP assignment of ISPs. Another factor can be that the RESIP services differ notably in their recruitment of RESIP nodes, which leads to the heterogeneous distribution of RESIP nodes across services.
Additionally, these observations highlight the quick evolution of the RESIP ecosystem and underscore the importance of continuous data collection, which is crucial for understanding the evolving landscape and addressing the security challenges in the RESIP ecosystem.

\subsection{RESIP Traffic}
\label{subsec:bs_traffic}
As detailed in \S\ref{subsec:method_bs_traffic}, 
We deployed RESIP nodes for the aforementioned 3 providers in both China and US, and captured their network traffic between June 03, 2022 and December 31, 2022. The resulting traffic logs (pcap files) are securely stored in our research servers, to which only authorized team members can have access. Upon these traffic logs, we generate statistical results as detailed below. Considering the majority of these traffic flows turn out to be RESIP traffic (i.e., traffic flows relayed by RESIPs), their statistical results can thus help us understand, for the first time, many characteristics of RESIP traffic, e.g., the magnitude, protocols, categories, etc. 

\finding{For the first time, a dataset of wild RESIP traffic has been collected, which is large-scale (3TB in size and 116M network flows), diverse (188K destination IP addresses), and representative (three RESIP services).}

\subject{Scale}. In total, we have captured 3,331.5 GBs of network traffic, comprising 116 million traffic flows (95,808,924 TCP flows, and 20,298,731 UDP flows). These traffic flows involve various traffic destinations of 188,015 IP addresses, 116,014 fully qualified domain names, and 50,526 apex domain names. 
/ronghong{ It is important to note here that, in our traffic destination address extraction, 95.31\% of the traffic can obtain domain names through SNI, which also shows that the use of DoH by Honeygain does not affect our statistical results. }
The statistics of traffic flows aggregated by RESIP providers and locations of RESIP nodes are listed in Table~\ref{tab:traffic_stats}. 

\begin{table}
    \centering
    \footnotesize
    \caption{The statistics of captured RESIP network traffic.}
    \label{tab:traffic_stats}
    \begin{threeparttable}
    \begin{tabular}{ccllll}
    \bottomrule
        Provider & Location & Size\tnote{1} & Flows & IPs & FQDNs\tnote{2}\\
        \midrule
                PacketStream & CN & 371 & 36.13M & 55K & 35K\\ 
        PacketStream & US & 2,735 & 69.76M& 125K & 78K\\ 
        IPRoyal & US & 57 & 2.06M & 23K & 11K\\ 
        Honeygain & US & 141 & 4.58M & 40K & 23K\\
        Honeygain & CN & 13 & 3.57M & 8K & 3K \\
        All & All & 3,332 & 116.10M& 188K& 116K\\
        \bottomrule
    \end{tabular}
        \begin{tablenotes}[flushleft]
        \item [1] Traffic volume in GBs.
        \item [2] Fully qualified domain names.
    \end{tablenotes}
    \end{threeparttable}
\end{table}

\subject{Protocols and Categories}. Here, we utilized the tool dpkt~\footnote{https://dpkt.readthedocs.io} to automatically decide the network protocols of each traffic flow. 
Among all traffic flows, 82.52\% are TCP flows, which account for 98.85\% traffic volume. The remaining UDP traffic flows are predominantly used for DNS queries/responses, accounting for over 99.99\% UDP traffic volume. A minuscule fraction of less than 0.01\% of the UDP flows is utilized for the SSDP (Simple Service Discovery Protocol) and NAT-PMP (NAT Port Mapping Protocol), which are employed for network services discovery and port mapping in NAT devices, respectively.

We then focus our following analysis on TCP flows. 
When dividing the TCP flows by their application layers, HTTPS traffic stands out and accounts for 92.17\% traffic flows and 98.63\% traffic volume, which is followed by HTTP traffic. The top application layer protocols ordered by their traffic volume are listed in Table~\ref{tab:traffic_services}. In addition to the dominating HTTP(S) traffic, it is surprising to see a non-negligible portion of the traffic is email traffic through SMTP. Further investigation turns out that all these traffic flows were used to relay email spamming activities, as illustrated in \S\ref{sec:sec_risks} when profiling malicious RESIP traffic.  



Since most traffic turns out to be web traffic (HTTP/HTTPS flows) towards 116,014 FQDNs,  we move to profile what kinds of websites have been relayed through RESIP nodes. 
We observe that web traffic flows show a long-tailed distribution across their FQDNs, and top 1,000 FQDNs make up 78.27\% web traffic flows and 88.11\% web traffic volume.
Therefore, we then looked into these top 1,000 FQDNs, aiming to decide their categories, and we refer to Similarweb for a complete list of website categories.
As a result, 219 FQDNs are considered Business and Consumer Services websites, which make up 12.12\% web flows and 23.2\% web traffic. What follows is the category of Shopping, which comprises 328 FQDNs. More categories can be found in Appendix~\ref{appendix:web_traffic_categories}. 

\begin{table}
    \centering
    \caption{Top application-layer services in RESIP traffic.}
    \footnotesize
    \label{tab:traffic_services}
    \begin{tabular}{ccll}
    \toprule
    Service & Typical Port & \% Flows & \% Volume\\
    \midrule
        HTTP & TCP 80 & 2.9\% & 0.56\%\\
        HTTPS & TCP  443& 92.17\% & 98.63\%\\
        HTTP(S) & TCP 80/443 & 95.06\% & 99.19\%\\
        SMTP & TCP 25/587 & 2.86\% & 0.53\%\\
        IMAP &  TCP 110/993 & 0.19\% & 0.03\%\\
        POP3 & TCP 143/995 & 0.01\% & 0.02\%\\
        Others & N/A & 1.06\% & 0.06\%\\
        Unknown & N/A & 0.82\% & 0.18\%\\
        \bottomrule
    \end{tabular}
\end{table}





\section{The Security Risks}
\label{sec:sec_risks}
Given relayed flows identified by our signature-based RESIP traffic classifier, we move to profile their security and privacy implications along with concrete cases and necessary statistical analysis. Compared with previous studies~\cite{mi2019resident, DBLP:conf/ndss/MiTLLQ021}, our analysis also reveals RESIP-relayed suspicious visits towards popular websites but at a much larger scale. More importantly, we have observed, for the first time, multiple previously unknown malicious traffic activities including visiting security-sensitive websites, Email spamming, and retrieving emails of compromised Email accounts.  


\subsection{Traffic Towards Popular Web Services}
\label{subsec:risks_popular_websites}
Among the relayed flows, most were found to be suspicious traffic towards popular online services, e.g., search engines, shopping websites, and online social networks. In total, the top 100 fully qualified domain names (FQDNs) account for 64.85\% of all the relayed traffic, and you may refer to Appendix~\ref{appendix:top_fqdns_of_resip} for the top 10 FQDNs that are most visited by RESIPs.
Also, flows towards the same popular online service often show up a pattern of traffic flooding. For instance, on date November 6, 2022, one RESIP node under our control had observed 78,405 flows towards \textit{oidc.idp.clogin.att.com}, a domain name associated with the authentication system used by US telecommunications company AT\&T. Also, it is unlikely that regular online users would relay their traffic towards popular online services through RESIPs, since traditional VPN services can provide even better anonymity and privacy for regular users at a much lower cost.  
For instance, it costs only \$6 per month for a subscription to a popular VPN service named NordVPN and such a subscription has no bandwidth limit. However, IPRoyal, a major RESIP service, charges users \$7 for every one GB traffic. 
All these factors render such RESIP flows suspicious, which echoes the observations by a previous study~\cite{mi2019resident} wherein the authors considered such traffic flows as blackhat SEO, advertisement fraud, and illicit data scraping, etc.  

\subsection{Traffic Towards Sensitive Websites}
\label{subsec:risk_sensitive_websites}


We observe that miscreants have been abusing RESIPs to relay their visiting traffic towards security-sensitive websites. This allows miscreants to masquerade their visits as benign from local residents and thus lower the possibility of being detected or even blocked. 
To the best of our knowledge, such suspicious activities of RESIPs have never been revealed in previous research studies~\cite{mi2019resident,DBLP:conf/ndss/MiTLLQ021, DBLP:conf/ccs/YangYMTGLZD22}.

\finding{We observe, for the first time, the adoption of RESIPs for masquerading suspicious visitors as local residents when  visiting sensitive web services operated by government/military/education agencies, web consoles of cyber physical systems,  and office automation systems, etc.}

\subject{Visiting government/military/education websites.} A website is typically considered as security sensitive when it is operated by government agencies, military organizations, or educational institutions. Such websites tend to be registered under special suffixes, e.g., a government website usually has the \textit{gov} in its suffix while it is \textit{edu} for education websites, and \textit{mil} for military ones.  In total,  20,713 traffic flows have been observed to visit one of 316 government websites, 9 military websites, and 560 education websites. Among these flows, 89.62\% were through HTTPs while the remaining flows were through HTTP. Among government websites, the most visited ones are subdomains of \textit{mfa.gov.hu}, a website offering consular service for the government of Hungary. Another example is \textit{reemployct.dol.ct.gov}, an unemployment service website operated by the Connecticut Department of Labor in the United States. Other examples include the official website of United States House of Representatives, the official website of The United States Social Security Administration, and the Treasury Check Verification System of the United States. However, since all flows are encrypted through HTTPS, we cannot learn the visit details.

Among military websites, two are operated by military agencies in China while the left 7 are in the United States. Also, they were exclusively observed by RESIP nodes located in the same country. To avoid potential security concerns, we wouldn't name them directly here. For instance, CN-mil-1 is a website running as an information management system for weapon purchase and bidding, and it is only reachable when visiting in China. Similarly, the 7 US military websites have also enforced location-based access control. However, the availability of RESIP services renders such kinds of location-based access control invalid. Regarding educational websites, websites of many top universities around the world have been observed, e.g., Duke University, the Ohio State University, and Columbia University. Still, we don't have access to the visit payload due to HTTPS.

\subject{Visiting sensitive websites running on unusual TCP ports}. Furthermore, many RESIP flows were found to be visits towards sensitive websites that appear to be not intended for public access.   In total, we have manually confirmed suspicious traffic towards 73 different security-sensitive websites. These websites belong to diverse categories, e.g., industrial control systems, airport control systems, watering control systems, remote desktops inside government agencies, etc.  Also,  all of these websites are located in China, and visits to these websites were relayed through RESIP nodes deployed in China. This echoes previous news reports~\cite{resip_in_attack} that residential proxies were used by hackers to masquerade themselves as residents in the USA when attacking critical web services in the USA. Apparently, attackers with their origins unclear to us, have abused residential proxies (BS nodes) in China and disguised their attacking traffic as originating from local residents. 

\subsubject{The categories of security-sensitive Chinese websites}. Despite a small scale of 73 different websites, these websites are of diverse categories. From the perspective of functionalities, 27 can be considered as web consoles to control cyber-physical systems (CPSes). Interesting and highly sensitive  CPSes include water systems, power plants (hydropower, solar power, and thermal power), power grids, airports, railways, flood monitoring systems, and environmental monitoring systems.

Another 24 websites are office automation (OA) systems, and most of them are either operated by government agencies or military agencies. What is even more concerning is that 3 out of these OA systems failed to enforce any access control despite storing datasets that either contain personally identifiable items or are critical to the respective organization.  In addition to CPS and OA,  another 3 websites are confirmed to be used for managing network devices (e.g., routers), and another one is used to manage cyberspace threat intelligence. For most of the remaining websites, we cannot decide their categories from the elements on the login pages to which we got redirected when manually visiting these websites. We also made efforts to decide the types of organizations to which these websites belong. It turns out that 31\% are operated by governments, 18\% are by military agencies, and another 21\% are by other organizations in the public sector, e.g., hospitals, education agencies, railway management companies. 

\subsubject{Access control for visits inside and outside China}. To investigate why visits to these websites were only observed on  RESIPs nodes deployed in China, we manually visited these websites from multiple locations inside and outside China and found that only 4 websites enforce access restrictions for visits from regions out of China. We thus believe the intention of relaying traffic through China RESIPS is to masquerade the visits as ones from benign local residents, so as to lower the chance of getting blocked or alarmed.

Another pattern of these sensitive websites is that none of them have domain names and their web servers were deployed on uncommon TCP ports, likely in an attempt to lower the visibility of their existence. Therefore,  to visit them, you must know the specific IP address along with the port number. Attackers must have got a list of such sensitive websites from other channels (e.g., large-scale port scanning), before relaying visiting traffic through RESIP nodes. 
To further profile how covert these websites are and how the initiators of these RESIP visits learn the addresses of these websites, we queried the Censys Universal Internet DataSet~\footnote{https://search.censys.io/data} which scans and indexes all the network services running on publicly accessible hosts. It turns out that only 39 out of the 73 websites have been indexed by Censys, which further highlights the stealthiness and suspiciousness of these RESIP visits.

\subsection{Suspicious or Malicious Email Traffic}
\label{subsec:risk_email_delivery}

\finding{For the first time, we observe that RESIPs are extensively abused in Email spamming activities, which gives spam operators a global reach of exit nodes and thus significantly undermines the effectiveness of anti-spam endeavors that are built upon blocklists of IP addresses or autonomous systems.}

\subject{Email spamming activities}. In addition to suspicious email retrieval activities, a much larger volume of email spamming traffic has been observed. 
Among RESIP traffic flows, 806,851 are found to be SMTP flows with the remote destination TCP port being 25, which means RESIPs are used to relay traffic of sending emails between SMTP servers. These SMTP connections involve a traffic volume of 6.33GB, among which, PacketStream accounts for 83.60\%, and IProyal has contributed all the left (16.40\%), while we didn't observe SMTP flows for Honeygain. From these SMTP flows,  we have successfully extracted 107,764 SMTP clients (uniquely identified by their FQDNs), 464,664 sender email addresses, 2,289,945 receiver email addresses, and 549,424 email messages, and 53 SMTP servers. 

\subsubject{Messages and templates of spam emails}. To further investigate what emails have been sent, we first manually studied a sampled set of email messages, all of which have been confirmed as spam emails along with a set of spamming templates identified.  Given these spamming templates, we built up a simple signature-based scanner, scanned the left email messages to decide if they belong to known spam templates.
Through this process, we conclude that 100\% of email messages are spam emails, and these spam emails were composed of 179 unique templates. Examples of spam examples can be found in Appendix~\ref{appendix:spam_email_templates}. These spam messages can be further divided into 2 spam categories: advertisement, and malware distribution.  
More stats on these spam categories can be found in Table~\ref{tab:spam_email_category} of the Appendices. Through looking into these spam emails and templates, we have also observed several commonly used evasion techniques including text paraphrasing, the use of homoglyphs, and the adoption of complicated encoding schemes, etc. More details can be found in Appendix~\ref{appendix:email_spam_evasion}. 

%


\subsubject{Recipients and senders of spam emails}. Also, a large number of Email users have been targeted by the observed spamming activities. Specifically, 2,289,945 unique email recipients have been observed in total, which are registered under 53 different SMTP servers, e.g., \textit{mail.protection.outlook.com} and \textit{mx.gmail.com}. 
We also profile the extent to which an email recipient has been targeted by email spamming. It turns out that 77.02\% spam email recipients have received only one spam email, while 100\% have received five or fewer. However, this can only serve as the lower-bound estimate since the RESIP nodes we deployed account for just a very small portion of all the RESIPs, and the email spam traffic observed on these nodes should be considered the tip of the iceberg for all the email spam traffic relayed through RESIPs.
Along with email messages and email recipients, spammers have masqueraded themselves as 464,664 email senders registered under 820 FQDNs. 

\subsubject{Delivery of spam emails}. Among all the 549K email messages, 104K have been successfully delivered through RESIPs towards 940K Email recipients, while the left failed for various factors elaborated in Appendix~\ref{appendix:failures_spam_email}.  Particularly, among all email delivery failures, 30.19\% are due to that the involved RESIP IP hit IP blocklists adopted by the SMTP server. And the adoption of IP blocklists by SMTP servers also explains why spamming operators choose to relay their traffic through globally distributed RESIPs. For instance, Gmail servers will assign ISPs with various spamming reputation scores, and a low reputation will make emails sent from the ISP be considered suspicious or even get blocked. Therefore, by leveraging RESIPs,  spamming operators can quickly switch to a new exit node and thus originate their spam traffic from a large number of different ISPs, which can likely get them a better reputation when sending new spam emails. 

\subject{Suspicious Email retrieving activities}. 
In addition to the visits towards sensitive websites, we have also observed a bunch of email retrieval flows. In total, 86,872 distinct flows were identified to visit 641 distinct email servers, e.g., \textit{imap.orange.fr}, \textit{imap.163.com}, \textit{pop.vid-eotron.ca}, and \textit{imap.uol.com.br}.  Also, 98.23\% flows are encrypted through IMAP over SSL or POP3 over SSL, while the remaining flows are plaintext traffic of either IMAP or POP3. 

Given plaintext traffic for email retrieval, we were allowed to analyze what emails had been retrieved from the server and for which accounts. Still, we followed a strict and ethical process wherein only statistical data was generated and presented. These 1,535 flows of plaintext email protocols involve 247 distinct Email accounts,  52.23\% of which have been successfully logged in with correct credentials. Also, among 193 retrieved emails, some were found to be security-sensitive ones that should not be exposed to the public. 
However, upon these results, we cannot decide whether these email traffic flows came from the authentic email account owners or from attackers with stolen email credentials. Instead, these traffic flows can be regarded as suspicious, considering the observed login failures as well as the low probability of having regular users using these expensive RESIPs. 

\subsection{Malicious Destination Addresses} 
Aligned with previous studies~\cite{mi2019resident}, we have also observed the misuse of RESIPs into masquerading visits towards malicious traffic destinations. 
Specifically, 
We queried several open threat intelligence exchanges  (OTX) to learn the maliciousness of the observed traffic destinations, namely, IPs, domain names, and URLs. These threat intelligence platforms include VirusTotal, IBM X-Force, and URLhaus, all of which are well-acknowledged in the cybersecurity community. Leveraging historical threat datasets of these platforms, we are allowed to profile the extent to which the relayed traffic flows are towards malicious traffic destinations as well as the category distribution of malicious traffic flows.

Given traffic destinations of  188,015  IP addresses, 116,014 FQDNs, and 50,526 apex domain names, we queried each OTX for the threat reports if any.
Overall, 15.27\% network addresses have been alerted as malicious by one or more OTXs. A closer look into the alarms turns out that most (6.8\%) were alarmed due to activities of \textit{anonymization services}, which is followed by \textit{botnet command and control servers} (6.69\%) and \textit{malware distribution} (5.22\%). More detailed threat statistics can be found in Appendix~\ref{appendix:malicious_usage_resip}. 






\subsection{Responsible Disclosure}
\label{subsec:disclosure}
Given aforementioned security risks, we have conducted responsible disclosure to not only administrators of sensitive websites visited by RESIPs, but also the involved RESIP services. Below, we provide more details about the process and the current results.

\subject{Responsible disclosure to administrators of sensitive websites}. As suspicious RESIP visits indicate potential security risks for the involved sensitive websites running on unusual TCP ports, we conducted responsible disclosure through sending disclosure emails to their contact Email addresses. To achieve this, contact Email addresses were manually collected at our best efforts by visiting these sensitive websites and searching the Internet, which led to the identification of contact emails for 33 out of 73 sensitive websites. Then, a disclosure email was sent to each contact Email address up to three times with a resending interval being one week. In each disclosure email, we specify the address of the sensitive website, the RESIP visits as observed on our RESIP nodes, along with a background introduction to RESIPs. 
By this writing, we have completed this disclosure process but have yet to receive any concrete response.  

\subject{Responsible disclosure to RESIP services.} We have also disclosed all the aforementioned security risks to the respective RESIP services through sending disclosure emails.  By this writing, we have received replies from IPRoyal and are working with its technical team to locate the specific abuse incidents. For the other two RESIP services, we haven't received any concrete response.

\section{RESIP Traffic Classification}
\label{sec:defense}

\begin{table}
    \footnotesize
    \centering
        \caption{The ground truth for RESIP traffic classification.}
    \label{tab:datasets_groundtruth}
    \begin{tabular}{cccc}
    \toprule
       Class  & Group & Flows  & Size (MB)\\
       \midrule
        \multirow{4}{*}{Relayed Flows} &   PacketStream  & 2,000 & 161.92\\
            & Honeygain & 2,000 & 74.33\\
            & IPRoyal & 2,000 & 791.33\\
            & Total & 6,000 & 1,027.58\\
            \hline
           \multirow{4}{*}{Tunnel Flows} &   PacketStream  & 2,000 & 88.29\\
            & Honeygain & 2,000 & 91.45\\
            & IPRoyal & 2,000 & 762.08\\
            & Total & 6,000 & 941.82\\
             \hline
            \multirow{2}{*}{Others} &  
                 VNAT~\cite{jorgensen2022extensible} & 1379  &418.20\\
         &ISCXVPN2016~\cite{draper2016characterization} & 3621 & 3082.19\\
         \bottomrule
    \end{tabular}
\end{table}


To defend against the non-negligible security risks of RESIP traffic, one promising option is to deploy a RESIP traffic detector at various vantage points, e.g., a local device, the network gateway of an organization, or a router operated by an ISP. Regardless of the vantage points, RESIP traffic will be mixed with non-RESIP flows that are both large-scale and of diverse categories, which requires the detector to be able to distinguish RESIP traffic from non-RESIP traffic. Also, the RESIP traffic detector should be generalizable to unknown RESIP services regardless of their protocols and implementation. All these factors render the aforementioned signature-based RESIP traffic classifier ineffective. Therefore,  we further pursue this task through the combination of machine learning, a set of robust features inherent in each traffic flow, and a large-scale ground truth dataset consisting of real-world relayed flows, tunnel flows, and non-RESIP traffic. 


\subject{Groundtruth}. We abstract the RESIP traffic detector as two binary classification tasks. One is to decide whether a given flow is a relayed flow or not, and the other is to classify whether a given flow is a tunnel flow or not. 
Table~\ref{tab:datasets_groundtruth} presents the ground truth datasets composed for both binary classification tasks.  Leveraging the signature-based RESIP traffic classifier (\ref{subsec:method_traffic_analyzer}), we have successfully sampled out 6,000 relayed flows and 6,000 tunnel flows in total covering all three RESIP services under our study. Furthermore, two publicly available network traffic datasets, namely, VNAT~\cite{jorgensen2022extensible} and ISCXVPN2016\cite{draper2016characterization}, have been adopted as the non-RESIP traffic flows. Among these two complementary datasets, ISCXVPN2016 has been well adopted in many previous studies for encrypted traffic classification~\cite{wang2017end, wang2017hast, he2020pert}, which consists of diverse traffic flows that are either VPN-tunnelled or not. Similarly, VNAT is a newly released dataset of traffic flows that are generated by 10 diverse Internet applications. When building up the classifiers for relayed flows, both tunnel flows and these two complementary datasets will be used as negative samples. On the other hand, when building up the classifiers for tunnel flows, the negative samples consist of the relayed flows and these two complementary datasets. 

\begin{table}
    \centering
        \caption{The performance of various classification models for RESIP relayed flow detection. 
        }
    \footnotesize
    \label{tab:model_performance_relayed_flow}
    \begin{threeparttable}
        \scriptsize 
        \begin{tabular}{cccccccc}
        \toprule
           Model &  $N_\text{up}$ & $N_\text{down}$ & $N_\text{all}$ & Precision & Recall & F1-Score & FPR\tnote{1}\\
           \midrule
             RF-BERT & NA & NA & 5 & 0.9287 & \textbf{0.9304} & \textbf{0.9296} & 0.0396 \\
            RF-RF  & 64 & 64 & 64 &  \textbf{0.9620} &  0.8880  & 0.9235 & \textbf{0.0191} \\
            RF-RF & 8 & 8 & 8 &  0.9455 & 0.8716 & 0.9071 & 0.0279 \\
            RF-RF & 4 & 4 & 4 & 0.9432 & 0.8741 & 0.9073 & 0.0293\\
            RF-RF & 4 & 0~\tnote{2} & 4 & 0.9338 & 0.8513 & 0.8907 & 0.0321\\
            \bottomrule
        \end{tabular}
        \begin{tablenotes}
            \item [1] FPR stands for the false positive rate.
            \item [2] Zero denotes the classifier doesn't require the availability of any specific number of packets in the respective direction (downstream or upstream).
        \end{tablenotes}
    \end{threeparttable}
\end{table}

Given the ground truth, for both classification tasks, We explore two machine learning paradigms. One is to fine-tune a pre-trained transformer-based model which allows us to automatically encode a traffic flow into a generic representation. The other is to manually design features and utilize classic classification algorithms to build up the classification models. Also, across classification models, 80\%  groundtruth are used for training and validation, while 20\% are held out for testing.  We then elaborate on both efforts below. 




\subject{Classifying RESIP flows through fine-tuning ET-BERT}. 
Here, we consider the classification of relayed flows and tunnel flows as the downstream tasks of fine-tuning ET-BERT, a BERT-like pre-trained model for generic network traffic representation (\S\ref{sec:background}). As listed in Table~\ref{tab:model_performance_relayed_flow}, the resulting DNN model, namely, RF-BERT, for relayed flow classification, has achieved a precision of 92.87\%, a recall of 93.04\%, and a low false positive rate of 3.96\%, while the resulting model for tunnel flow classification, namely,  TF-BERT has achieved a slightly lower but still comparable performance (Table~\ref{tab:model_performance_tunnel_flow}). 
One thing to note, our transformer-based classification models take as input only the first 5 packets of a given traffic flow, which allows the early detection of ongoing RESIP traffic flows.  
However, despite achieving a good prediction performance for both classification tasks, DNN models are known for their limited explainability, we thus move to explore whether the combination of manually crafted features and classic classification algorithms can achieve comparable performance.

\begin{table}
    \centering
        \caption{The performance of various classification models for RESIP tunnel flow detection.}
    \footnotesize
    \label{tab:model_performance_tunnel_flow}
    \begin{threeparttable}
     \scriptsize 
        \begin{tabular}{cccccccc}
        \toprule
           Model &  $N_\text{up}$ & $N_\text{down}$ & $N_\text{all}$ & Precision & Recall & F1-Score & FPR\tnote{1}\\
           \midrule
             TF-BERT & NA & NA & 5 & 0.9036 & 0.8891 & 0.8963 & 0.0535 \\
            TF-RF  & 64 & 64 & 64 &  \textbf{0.9571} &  0.9139  & 0.9350 & \textbf{0.0223} \\
            TF-RF & 8 & 8 & 8 &  0.9563 & 0.9012 & 0.9280 & 0.0229 \\
            TF-RF & 4 & 4 & 4 & 0.9515 & \textbf{0.9202} & \textbf{0.9356} & 0.0261\\
            TF-RF & 4 & 0\tnote{2} & 4 & 0.9487 & 0.9194 & 0.9338 & 0.0261\\
            \bottomrule
        \end{tabular}
        \begin{tablenotes}
             \item [1] FPR stands for the false positive rate.
            \item [2] Zero denotes the classifier doesn't require the availability of any specific number of packets in the respective direction (downstream or upstream).
        \end{tablenotes}
    \end{threeparttable}
\end{table}


\subject{Classifying RESIP flows with expert-crafted features and classic ML algorithms}. Next, we elaborate the features we have crafted, as well as the evaluation results achieved by classic classification algorithms. 

\subsubject{Manual feature engineering}. 
Before presenting feature groups in detail, we need to define three variables. One is $N_{\textsf{up}}$,  the maximum number of upstream packets to consider when composing the features, and the 2nd one is $N_{\textsf{down}}$ which refers to the maximum number of downstream packets to ingest, and the last is $N_{\textsf{all}}$, the maximum number of bi-directional packets to consider when composing features. As you can see, the larger these variables are, the more packets the classification model will ingest, and likely the more time it takes to wait for the packets before giving the classification results. 
We will build classification models with different $N_{\textsf{up}}$,  $N_{\textsf{down}}$, and $N_{\textsf{all}}$ so as to evaluate the trade-off between classification timeliness and classification effectiveness, i.e., how timely we can capture a RESIP flow with high confidence. By default, we set up all three variables as 64.


First of all, we observe that the tunnel flow tends to have a much higher upstream traffic volume than its downstream traffic, resulting in the size of outgoing packets being much larger than the size of received packets, which is consistent with observations of previous works~\cite{DBLP:conf/ndss/MiTLLQ021}. Therefore, a set of features are defined to profile the ratio of upstream traffic over all the traffic for each given traffic flow.
Then, we adopt a set of temporal or statistical features as proposed in previous works~\cite{draper2016characterization,he2020pert}, e.g., features to profile the inter-packet arrival timing and features to profile the flow throughput and packet size. More details for the full list of features can be found in Appendix~\ref{appendix:resip_features}. In total, a set of 186 features have been crafted for $N_{\textsf{up}}$ = $N_{\textsf{down}}$ = $N_{\textsf{all}} = 64$. All these features are normalized before training. One thing to highlight is that different from the automatic feature engineering through ET-BERT, these hand-crafted features don't depend on any packet payload, and are therefore robust to traffic encryption variations.

\subsubject{Training and evaluation}. During training and evaluation, multiple classic classification algorithms have been explored, which include decision tree, SVM, and random forest. We also experimented with various feature engineering settings, i.e., different combinations of $N_{\textsf{up}}$, $N_{\textsf{down}}$, and $N_{\textsf{all}}$, so as to understand how many packets are necessary for an effective classification model. As a result,  the random forest (RF) algorithm has consistently achieved the best performance and is thus chosen as the default algorithm for both classification tasks. Table~\ref{tab:model_performance_relayed_flow} presents the performance results of these random forest classifiers under the model name of \textit{RF-RF} for relayed flow classification, while classification models for tunnel flow classification are illustrated in Table~\ref{tab:model_performance_tunnel_flow} under the name of \textit{TF-RF}. 

As you can see, 
our feature-based models can still achieve a good detection performance even if only using the first 4 upstream packets, which suggests our classification models can be deployed to generate \textit{early} alarms for ongoing traffic flows. For instance, given features extracted from up to the first 4 upstream packets, our feature-based models can achieve a recall of 85.13\% and a precision of 93.38\%  for relayed RESIP flow classification, and a recall of 91.94\% and a precision of 94.87\% for tunnel RESIP flow classification. Also, as learned from our groundtruth dataset, 94.15\% relayed flows have 4 or more upstream packets, while 70.10\% have 8 or more. On the other hand, 99.93\% tunnel flows will not terminate after having 4 upstream packets sent out, while it is 99.88\% for 8 upstream packets. These results suggest that our detection models, especially the ones ingesting only the first 4 upstream packets, can alert and block \textit{most} RESIP flows before they have been closed.

Then, comparing the two classification options, the classic option can consistently achieve a lower false positive rate and a higher precision than the BERT option, along with a comparable performance in terms of recall.

\begin{table}
    \centering
        \caption{Throughput of RESIP traffic classification models. 
        }
    \footnotesize
    \label{tab:model_throughput}
    \begin{threeparttable}
        \scriptsize 
        \begin{tabular}{cccccc}
        \toprule
           Model &  $N_\text{up}$ & $N_\text{down}$ & $N_\text{all}$ & Task &  Throughput (flows/sec) \\
           \midrule
            RF-BERT & NA & NA & 5 & Relayed Flow&   98.7\\
            TF-BERT & NA & NA & 5 &  Tunneled Flow & 114.9 \\
            RF-RF & 64 & 64 & 64 & Relayed Flow &  298.3\\
            TF-RF & 64 & 64 & 64 & Tunneled Flow & 332.1\\

            \bottomrule
        \end{tabular}
    \end{threeparttable}
\end{table}

\subject{Throughput.}
Furthermore, we evaluate the throughput of models on an Ubuntu 21.10 system equipped with a 2.40GHz CPU and two GeForce RTX 3090 GPUs. For the BERT models, we run them on the the GPU , while the classic models are tested on the CPU. The throughput of these models are listed in Table~\ref{tab:model_throughput}. For instance, the RF-RF model achieves a throughput of 298 flows per second and per process, while it is 114 flows per second per process for the BERT counterpart. These throughput results underscore efficiency of proposed models as well as their suitability for meeting detection demands in various scenarios.

\finding{Regarding RESIP flow classification, random forest models with well-crafted features can achieve not only better performance but also higher throughput, when compared with BERT counterparts.}

\subject{Feature importance.} We have also measured the importance of each crafted feature using the metric of mean impurity decrease. In a nutshell, across feature engineering settings,  features describing packet arrival time for upstream flows and packet size for upstream flows are among the most important features for relayed flow classification, while, for tunnel flow classification, they are features profiling the upstream traffic ratio and features profiling packet arrival time for upstream flows. The top 10 features ordered by their importance for their respective classification models can be found in Appendix~\ref{appendix:feature_importance}.








\begin{table}
    \centering
        \caption{The influence of the VNAT\&ISCXVPN2016 traffic datasets.
        }
    \footnotesize
    \label{tab:model_influence_of_other_dataset}
    \begin{threeparttable}
        \scriptsize 
        \begin{tabular}{cccccc}
        \toprule
           Model & Whether include &  Precision & Recall & F1-Score & FP \\
           \midrule
             \multirow{2}{*}{RF-BERT} 
                & Yes & 0.9287 & 0.9304 & 0.9296 & 0.0396\\
                & No & 0.9055 & 0.9121 & 0.9088 & 0.0930\\
            \multirow{2}{*}{TF-BERT} 
                & Yes & 0.9036 & 0.8891 & 0.8963 & 0.0535\\
                & No & 0.8967 & 0.8694 & 0.8828 & 0.0993\\
            \multirow{2}{*}{RF-RF~\tnote{1}} 
                & Yes &  0.9620 & 0.8880 & 0.9235 & 0.0191\\
                & No & 0.9336 & 0.9064  & 0.9198 & 0.0607\\
            \multirow{2}{*}{TF-RF~\tnote{2}} 
                & Yes & 0.9571 & 0.9139 & 0.9350 & 0.0223\\
                & No &  0.9645 & 0.9330 & 0.9485 & 0.0324\\

            \bottomrule
        \end{tabular}
        \begin{tablenotes}
            \item [1] RF-RF refers to the best-performing random forest model that ingests 64 packets.
            \item [2] TF-TF refers to the best-performing random forest model that ingests 64 packets.
        \end{tablenotes}
    \end{threeparttable}
\end{table}

\subject{The influence of the VNAT \& ISCXVPN2016 traffic datasets.}
To investigate the effect of the VNAT \& ISCXVPN2016 dataset on our model's performance, we embarked on a series of ablation experiments. Upon excluding the VNAT \& ISCXVPN2016 datasets, we utilized only the datasets from three providers (Packetstream, Honeygain, and Iproyal) for both training and testing purposes. As shown in Table~\ref{tab:model_influence_of_other_dataset}, the VNAT \& ISCXVPN2016 datasets obviously benefit the model performance with regards to lowering the false positive rate. For instance, without  VNAT \& ISCXVPN2016 datasets, the random forest model for relayed flow classification (RF-RF) presents a decline in precision from 96.20\% to 93.36\%, and an escalation in the false positive rate from 1.91\% to 6.07\%. These substantial shifts underscore the pivotal role of the VNAT\& ISCXVPN2016 datasets in enhancing the model's generalization capabilities and minimizing false positives.


\section{Discussion}
\label{sec:discuss}
\subject{Mitigation recommendations}.
Considering the security and privacy risks of RESIPs, we move to discuss several promising defense techniques. First of all, organizations or ISPs may have the motivation to detect or even block relaying traffic through residential devices in their networks, which can be achieved by leveraging the classifiers we have built up to detect both tunnel flows and relayed flows. Also, similarly, traffic destinations may want to distinguish flows relayed through RESIPs from others, and our classifier for relayed flows can benefit this effort. 

Besides, when deploying a RESIP node,  none of the RESIP services under our study require authorization and confirmation from the authentic device owner. 
This gives attackers the opportunity to monetize compromised devices by deploying RESIP nodes~\cite{DBLP:conf/ndss/MiTLLQ021}. To request authorization from a device owner, a promising solution is to request authorization by prompting a confirmation dialog enforced by trusted execution environments (TEE), which tends to be increasingly available in servers and consumer devices~\cite{apple_enclave,android_protected_dialog}.
Specifically, when deploying a RESIP node on a device, a TEE-backed confirmation prompt is displayed on a trusted screen, and the device owner is required to give confirmation by pressing a physical button equipped with the device. And the confirmation result will be signed by the TEE using securely stored key materials before sending it to the RESIP server to verify. 
And the RESIP node will not be set up until the RESIP server has 
received and verified the signed confirmation results. A similar remote authorization protocol has been recently proposed and evaluated for the Android platform~\cite{imran2022sara}.

\subject{Data and code release}. 
To foster future research in relevant domains, we plan to release the source code of our RESIP traffic collector and the scripts for training and evaluating our RESIP traffic classifiers. As the RESIP traffic dataset involves both plain-text flows and flows towards vulnerable websites,  it will be available upon request and necessary background vetting. 
%

%
\section{Conclusion}
\label{sec:conclusion}
In this study, a RESIP traffic collector is proposed and built up to collect realistic RESIP traffic. Given the collected RESIP traffic, a RESIP traffic analyzer has been applied, through which, a set of novel and concerning security findings have been distilled regarding the malicious usage of RESIPs. Furthermore, to help mitigate the security risks of RESIPs, multiple robust machine learning classifiers have been built up so as to classify RESIP traffic in a timely and effective manner. 

\bibliographystyle{plain}
\bibliography{ref}
\appendix

\section{The Categories of Web Traffic Relayed by RESIPs}
\label{appendix:web_traffic_categories}

Table~\ref{tab:cate_web_traffic} presents the top categories of web traffic as relayed by RESIP nodes under our control. 
\begin{table}[h]
    \centering
    \footnotesize
    \caption{Top categories of web traffic relayed by RESIPs.}
    \label{tab:cate_web_traffic}
    \begin{threeparttable}
    \begin{tabular}{cccc}
        \toprule
        Category & \% FQDNs & \% Flows & \% Traffic \\
        \midrule
        eCommerce \& Shopping & 32.8\% & 29.25\% & 20.44\%\\
                B\&C\tnote{1}  & 21.9\% & 12.12\% & 23.2\%\\
        CET\tnote{2} & 5.4\% & 8.56\% & 13.57\% \\
        Arts \& Entertainment & 3.8\% & 1.93\% & 7.61\%\\
        Search Engine & 2.7\% & 2.89\% & 3.82\% \\
        \bottomrule
    \end{tabular}
     \begin{tablenotes}[flushleft]
        \item [1]  Business and Consumer Services
        \item [2] Computers Electronics and Technology 
    \end{tablenotes}
    \end{threeparttable}
\end{table}

\section{Suspicious RESIP Traffic Towards Popular Web Services}
\label{appendix:top_fqdns_of_resip}
Table~\ref{tab:top_fqdns} presents the top 10 FQDNs with most RESIP traffic observed. 

\begin{table}[h]
    \centering
    \caption{The top 10 FQDNs (fully qualified domain names) with most relayed traffic observed.}
    \label{tab:top_fqdns}
    \begin{tabular}{ccc}
        \toprule
       FQDN  & Flows & Volume \\
       \midrule
       googleapis.com & 2.39\% & 14.94\%\\
        www.instagram.com & 2.06\% & 6.12\% \\
        www.youtube.com & 1.29\% & 3.63\%  \\
        www.amazon.com & 1.25\% & 2.9\% \\
        i.instagram.com & 5.6\% & 2.67\% \\
        www.yeezysupply.com & 0.54\% & 1.85\%  \\
        www.google.com & 2.76\% & 1.78\%  \\
        m.youtube.com & 0.85\% & 1.32\%  \\
        www.gstatic.com & 0.28\% & 1.24\%  \\
        www.amazon.co.jp & 8.74\% & 1.21\%  \\
        \bottomrule
    \end{tabular}
\end{table}

\section{Categories and Templates of Spam Emails}
\label{appendix:spam_email_templates}
\begin{figure}[h]
    \centering
    \subfigure[
        A  spam email relayed by IPRoyal
        \label{fig:spam_template_1}
    ]{
        \includegraphics[width=.45\columnwidth]{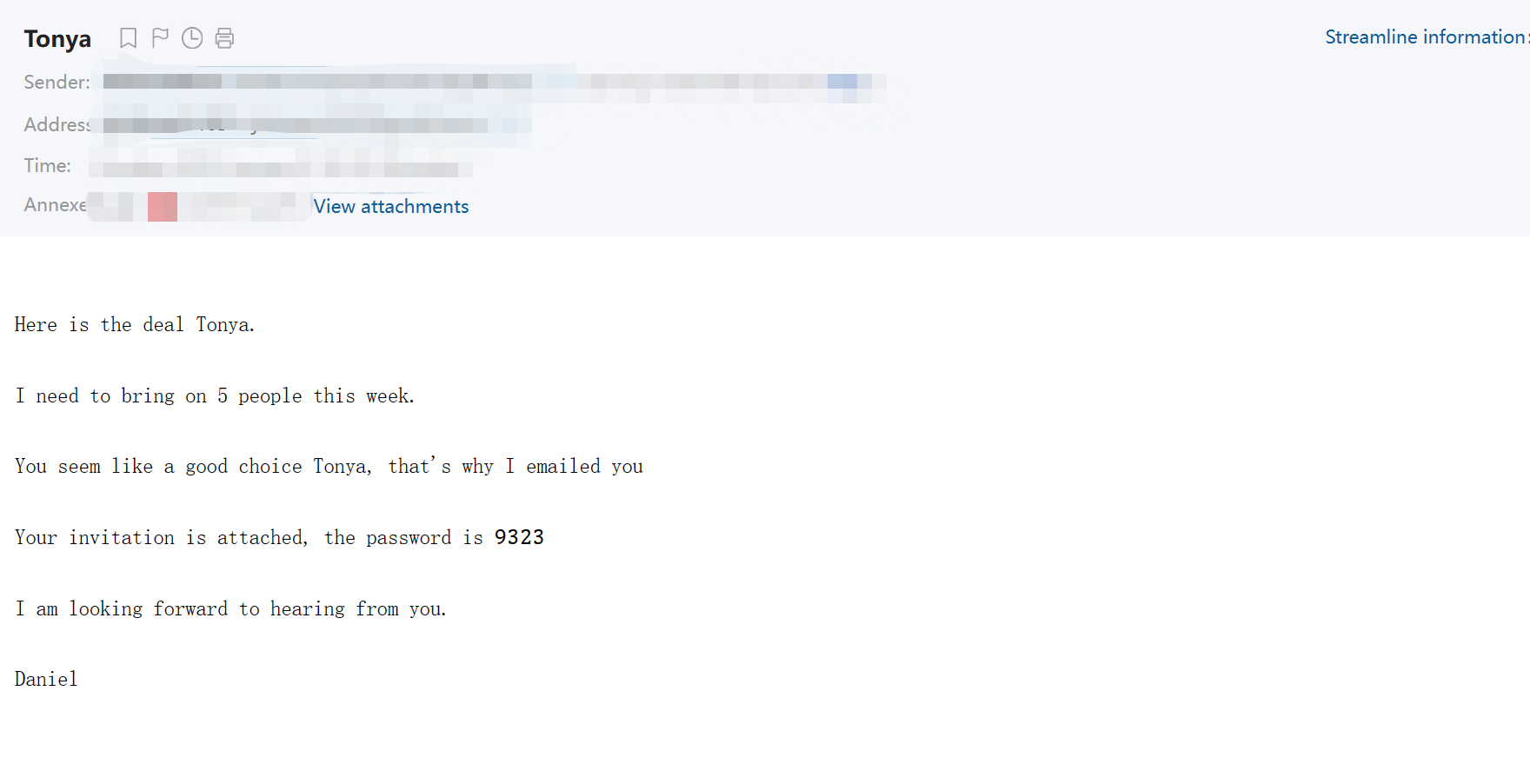}
    }
    \hfill
    \subfigure[
         A  spam email relayed by PacketStream
        \label{fig:spam_template_2}
        ]{
        \includegraphics[width=.45\columnwidth]{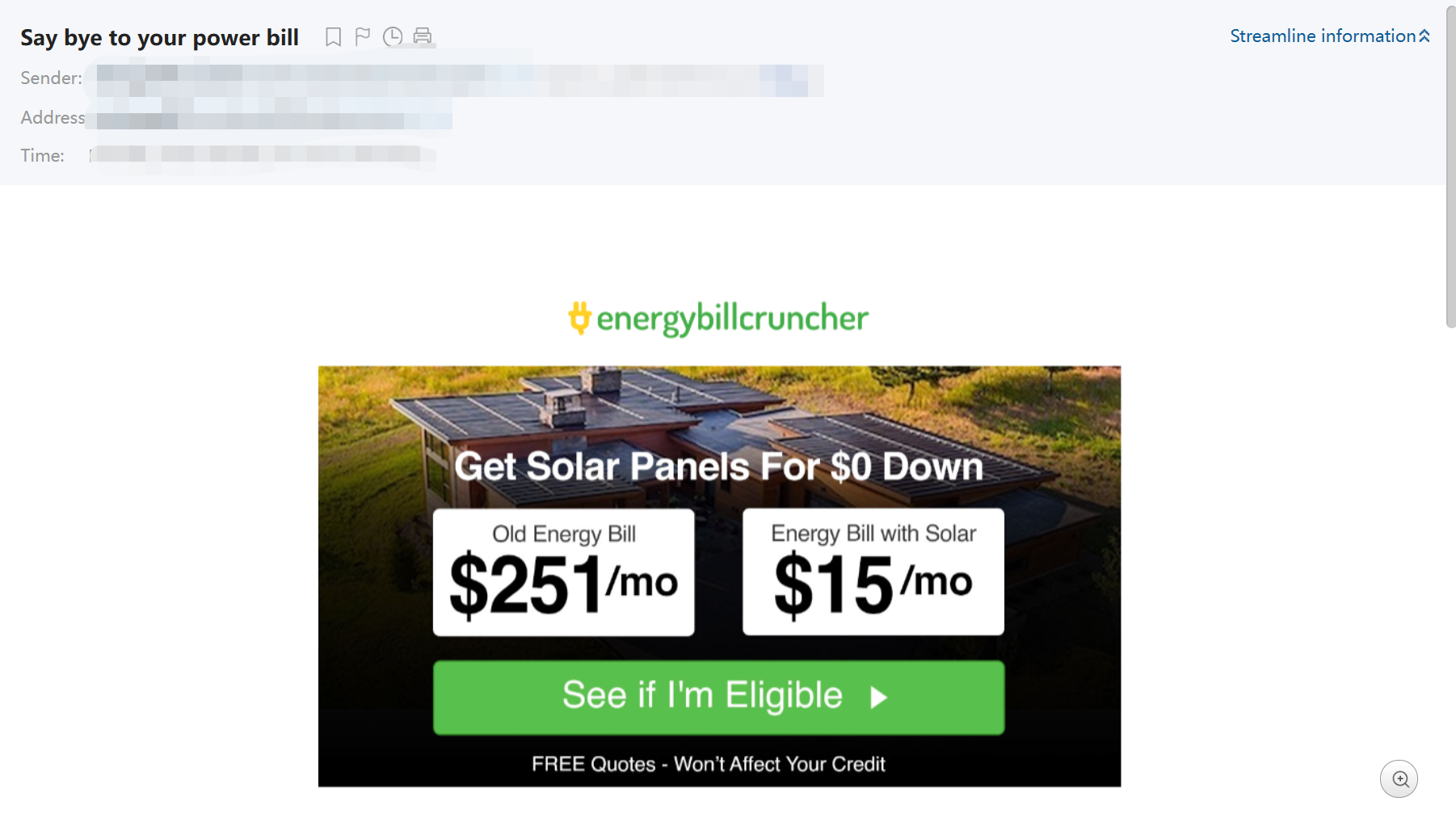}
    }
    \caption{Examples of spam emails.}
    \label{fig:email_spam_templates}
\end{figure}

Figure~\ref{fig:email_spam_templates} presents two templates of spam emails observed in RESIP traffic, while the distribution of spam Emails across their categories are listed in Table~\ref{tab:spam_email_category}. 

\begin{table}[h!]
    \centering
        \caption{The stats of spam emails from the perspective of categories.}
        \small
    \label{tab:spam_email_category}
    \begin{tabular}{cccc}
        \toprule
         Spam Category & Templates & \% Emails & \% Recipients \\
         \midrule
         Advertisement & 175 &  13.61\% & 58.71\% \\
         Malware Distribution & 4 & 86.39\% & 41.29\% \\
         \bottomrule
    \end{tabular}
\end{table}

\section{Evasion Techniques of Email Spamming}
\label{appendix:email_spam_evasion}
Among spam emails, one commonly used evasion technique is text paraphrasing. For instance, in a bunch of spam emails on job offer fraud, the text of \textit{I have attached your invitation} has been paraphrased to tens of different versions, e.g., \textit{Your invite is attached}, \textit{I have attached your secure invite}, and \textit{The invite is attached}. This evasion technique may effectively help spammers bypass naive blocklists of email messages. Another more advanced evasion technique utilizes complicated combinations of email encoding schemes, character encoding schemes, and homoglyphs. One typical combination as adopted by spammers is the combination of the UTF-8 character encoding standard and the  quoted-printable (QP) email encoding system. And the QP encoding is an encoding system 
commonly used to encode 8-bit data so that it can be transmitted over a 7-bit path. 


Here is a part of a spam email: ``\textit{the password is: =F0=9D=9F=BF=F0-=9D=9F=B9=F0=9D=9F=B8=\-F0=9D=9F=B9.=0D=0A   =0D=0AI am eager to hear back from you}''. Its true text is: \textit{the password
is: \texttt{9323}, I am eager to hear back from you}. In a nutshell, \texttt{9323} are Unicode characters defined to represent the respective decimal digits 9323 in the monospace font. Such kinds of Unicode characters belong to the Unicode block of mathematical alphanumeric symbols~\cite{math_symbols} which are designed as styled forms of letters and decimal digits for denoting different notions in mathematics. Taking \texttt{9} as an example, its unicode number is U+1D7FF, and its UTF-8 encoding is 0xF00x9D0x9F0xBF. Once encoded through QP, the resulting characters in the email message will be \textit{=F0=9D=9F=BF}. Unicode recommends that these characters should not be used in general text, however, they have been abused by spam operators to evade spam filtering.

\section{Spam Email Delivery}
\label{appendix:failures_spam_email}
 Leveraging the Email spamming traffic flows relayed by our RESIP nodes, a non-negligible portion of many have failed to deliver the emails for various factors. Specifically, 36.26\% flows got rejected by the SMTP server even before the client HELO command, and 40.04\% got closed by the server before the client told the server the recipients through the RCPT commands. Among the left SMTP flows which got the email messages successfully sent, 76.0\% got rejected along with various error messages, and only 24.0\% flows succeeded in delivering the emails, which involved 179 spam templates, 104,299 emails, and 939,679 recipients.

Among all the delivery failures, the causes can be summarized as the following groups, as revealed by the error messages from the SMTP servers. 
Example error messages from  SMTP servers include  ``\textit{Our system has detected that this message is likely unsolicited mail.}'',   ``\textit{Our system has detected an unusual rate of unsolicited mail originating from your IP address.}'', ``\textit{The mail server XX has been temporarily rate limited due to IP reputation.}'', and ``\textit{Please contact your Internet service provider since part of their network is on our block list}''. 
Among all email delivery failures, 30.19\% are due to the involved RESIP IP hitting one of the IP blocklists adopted by the SMTP server, 69.49\% failed because the email content triggered server-side spam filtering rules and 0.31\% due to message failing authentication checks. 

The adoption of IP blocklists by SMTP servers also explains why spamming operators choose to relay their traffic through globally distributed RESIPs. Particularly, Gmail servers will assign ISPs with various spamming reputation scores, and a low reputation will make emails sent from the ISP be considered suspicious or even get blocked. Therefore, by leveraging RESIPs,  spamming operators can quickly switch to a new exit node and thus originate their spam traffic from a large number of different ISPs, which can likely get them a better reputation when sending new spam emails. However, the abuse of RESIPs in spamming activities highlights another concerning issue that the reputation of a clean ISP could be heavily polluted when there were local RESIP nodes and these RESIP nodes got abused by spamming operators.

\section{RESIP Traffic Towards Malicious Destinations}
\label{appendix:malicious_usage_resip}
Table~\ref{tab:threat_stat} presents the threat statistics of traffic destinations observed in RESIP traffic. 

\begin{table}[h]
    \centering
    \footnotesize
        \caption{Threat stats for RESIP traffic destinations.}
    \label{tab:threat_stat}
    \begin{tabular}{cccc}
        \toprule
         OTX Platform & IPs &  Domains &  URLs \\
         \midrule
         VirusTotal& 0.68\% & 2.16\% & 0\\
         IBM X-Force & 14.88\% & 3.22\% & 4.5\%\\
         URLhaus & NA & 0.0034\% & 0\\
         All & 15.27\% & 3.24\% & 4.5\%\\
         \bottomrule
    \end{tabular}
\end{table}

\section{Manually Crafted Features for RESIP Traffic Classification}
\label{appendix:resip_features}

\begin{table}
    \centering
    \footnotesize
    \caption{The most important features ordered by their contribution to the decrease in the mean impurity for the respective random forest classifier. }
    \label{tab:top_features_of_flow_classification}
    \begin{threeparttable}
        \begin{tabular}{ccccc}
            \toprule
        No & $\textsf{RF-RF}_{8, 8, 8}$\tnote{1}& $\textsf{RF-RF}_{4, 0, 4}$\tnote{1} & $\textsf{TF-RF}_{8, 8, 8}$\tnote{1}& $\textsf{TF-RF}_{4, 0, 4}$\tnote{1} \\
        \midrule
           1  & $\text{PAT}_{up, 4, min}$\tnote{2} & $\text{PAT}_{up, 4, min}$ & $\text{UpRatio}_{8}$ & $\text{UpRatio}_{4}$ \\
           2  &
           $\text{UpRatio}_{8}$\tnote{2} & $\text{PS}_{up, 4, max}$\tnote{3} & $\text{PAT}_{up, 4, min}$ & $\text{PAT}_{up, 4, min}$\\
           3 &            $\text{PS}_{up, 8, max}$ & $\text{PS}_{all, 4, mean}$ & $\text{PS}_{down, 8, mean}$ & $\text{PS}_{up, 4, max}$ \\
           4 & $\text{PS}_{up, 4, max}$ & $\text{PS}_{up, 4, std}$ & $\text{PS}_{up, 8, max}$ & $\text{PS}_{all, 4, mean}$ \\
           5 & $\text{PS}_{all, 4, mean}$ & $\text{PS}_{up, 4, mean}$ & $\text{PS}_{all, 4, mean}$ & $\text{PS}_{all, 4, max}$ \\
           6 & $\text{PS}_{up, 4, mean}$ & $\text{PS}_{all, 4, std}$ & $\text{PS}_{all, 4, max}$ & $\text{PS}_{up, 4, std}$ \\
           7 & $\text{PS}_{up, 8, std}$ & $\text{BPS}_{down, 4}$ & $\text{PS}_{up, 8, std}$ & $\text{PS}_{down, 4, mean}$ \\
           8 & $\text{PS}_{all, 4, max}$ & $\text{PS}_{all, 4, max}$ & $\text{PS}_{up, 4, max}$ & $\text{PS}_{all, 4, std}$ \\
           9 & $\text{PS}_{down, 8, mean}$ & $\text{PPS}_{all, 4}$ \tnote{3} & $\text{PS}_{all, 4, std}$ & $\text{PS}_{up, 4, mean}$ \\
           10 & $\text{PS}_{up, 4, std}$ & $\text{BPS}_{all, 4}$~\tnote{3} & $\text{PS}_{up, 4, std}$ & $\text{PAT}_{up, 2, mean}$ \\
           \bottomrule
        \end{tabular}
        \begin{tablenotes}
            \item [1] $\textsf{RF-RF}_{8, 8, 8}$ denotes the random forest model for RESIP relayed flow classification by ingesting only the first 8 packets in upstream/downstream/bi-directional traffic, while $\textsf{RF-RF}_{4, 0, 4}$ considers only the first 4 packets in upstream and bi-directional traffic. Similarly, $\textsf{TF-RF}_{8, 8, 8}$ shares the same feature setting with $\textsf{RF-RF}_{8, 8, 8}$, but for tunnel flow classification. The same case applies to $\textsf{TF-RF}_{4, 0, 4}$.
            \item [2] $\text{PAT}_{up, 4, min}$ denotes the minimum value of the packet arrival time for the first 4 upstream packets. Similarly, $\text{UpRatio}_{8}$ denotes the ratio of upstream traffic over all the traffic by the 8 upstream packets.
            \item [3] $\text{PS}_{up, 4, max}$ refers to the maximum packet size for the first 4 upstream packets, while $\text{BPS}_{all, 4}$ denotes the throughput in bytes per second, and $\text{PPS}_{all, 4}$ denotes the throughput in packets per second, both for the first 4 packets of a flow.
        \end{tablenotes}
    \end{threeparttable}
\end{table}

Below, we detail all the features we have crafted for RESIP traffic classification. 
First of all, assume that $\textsf{UpLen}_{i}$ and $\textsf{DownLen}_{i}$ are the cumulative length in bytes of the given stream up to the $i\text{th}$ upload packet.
We define $\text{UpRatio}_{i} = \frac{\textsf{UpLen}_{i}}{\textsf{UpLen}_{i} + \textsf{DownLen}_{i}}$ to denote the ratio of upstream traffic over all the traffic by the $i\textsf{th}$ upstream (outgoing) packet. 
We then extract features for $i = \{2, 4, 8, \ldots, N_{\textsf{up}}\}$, an exponential sequence with a base of 2 and the largest value being $N_{\textsf{up}}$, to profile the ratios of upstream traffic. 
One thing to note, when the number of all upstream packets in the whole flow ($FP_{up}$) is smaller than $N_{\textsf{up}}$, the tailing $\text{UpRatio}$ features will be set as the last valid $\text{UpRatio}_{i}$.  For instance, when $FP_{up} = 29$ and $N_{\textsf{up}} = 64$, $\text{UpRatio}_{32}$ and $\text{UpRatio}_{64}$ will have the same value as $\text{UpRatio}_{16}$. This rule is also applied to all other features introduced as below, so as to generate same-sized feature vectors for flows of varying packet counts.

Then, we adopt a set of temporal or statistical features as proposed in previous works~\cite{draper2016characterization,he2020pert}. Among these features, one set is designed to profile the \textit{inter-packet arrival time}  for upstream traffic of a given flow. Given the first $i$ upstream packets of a flow, the inter-arrival time is profiled using four statistical metrics: the mean, the minimum, the maximum, and the standard deviation. Also, $i$ is configured to be one of $i = \{2, 4, 8, \ldots, N_{\textsf{up}}\}$,  Which gives us another set of 24 features when $N_{\textsf{up}} = 64$. Similarly, 48 more features are defined to capture inter-packet arrival time for downstream and bi-directional traffic. Besides, the flow throughput is profiled as both bytes per second and packets per second, for the first $i$ packets. Still, these throughput values are calculated separately for the upstream, downstream, and bi-directional packets. This generates another set of 36 features. 
Furthermore, the packet size is profiled, through the aforementioned four descriptive statistical metrics, still, for the first $i$ packets in downstream, upstream, and bi-directional traffic, which leads to another set of 72 features.

\section{Feature Importance in RESIP Traffic Classification}
\label{appendix:feature_importance}
As detailed in \S\ref{sec:defense} regarding how to detect both RESIP relayed flows and RESIP tunnel flows, multiple groups of features have been designed to build up the feature-based classification models. Table~\ref{tab:top_features_of_flow_classification} lists the top 10 most important features that are ordered by their contribution to the respective random forest model in terms of decreasing the mean impurity across trees in the forest.
\end{document}